\documentclass[11pt,twoside]{article}
\usepackage{asp2010}
\usepackage{graphicx}

\resetcounters

\newcommand{\half}{\frac{1}{2}}

\begin{document}

\title{A numerical model of resistive generation of intergalactic magnetic field at cosmic dawn}
\author{Francesco Miniati$^1$, and A. R. Bell$^2$
\affil{$^1$Physics Department, Wolfgang-Pauli-Strasse 27,
ETH-Z\"urich, CH-8093, Z\"urich, Switzerland}
\affil{$^2$Clarendon Laboratory, University of Oxford, Parks Road, Oxford UK OX1 3PU}}

\begin{abstract}
\citet{Miniati2011ApJ...729...73M} proposed a mechanism for the generation of magnetic 
seeds that is based the finite resistivity of the low temperature IGM in the high redshift universe.
In this model, cosmic-ray protons generated by the first generation of galaxies,
escape into the intergalactic medium carrying an electric current that induces 
return currents, $j_t$, and associated electric fields, $\vec E=\eta\vec j_t$ there.
Because the resistivity, $\eta$, depends on the IGM temperature, which 
is highly inhomogeneous due to adiabatic contraction and shocks produced by structure 
formation, a non-vanishing curl of the electric field exists which sustains the growth of magnetic field.
In this contribution we have developed an approximate numerical model for this process
by implementing the source terms of the resistive mechanism in the cosmological code {\tt CHARM}.
Our numerical estimates substantiate the earlier analysis in~\cite{Miniati2011ApJ...729...73M}
which found magnetic seeds between 10$^{-18}$ and 10$^{-16}$ Gauss throughout 
cosmic space at redshift $z\sim 6$, consistent with conservative estimates of magnetic 
fields in voids at $z\sim 0$ from recent gamma-ray experiments.
\end{abstract}

\section{Introduction}
Magnetic fields are an intrinsic part of astrophysical plasmas, though in 
many respects their origin and role is still subject of intense investigation.
Recent studies combining gamma-ray measurements at different energies
indicate the existence of magnetic fields in cosmic 
voids~\citep{neronovandvovk10,Taylor2011A&A...529A.144T,Tavecchio2011MNRAS.414.3566T}. If confirmed
(see e.g.~\cite{2011arXiv1106.5494B}),
the origin of such magnetic fields poses a non trivial challenge for astrophysical 
scenarios: the mechanisms connected to the Weibel instability~\citep{schsh03,msk06} 
and the Biermann's battery at shocks~\citep{kcor97} generate magnetic fields inside collapsed structures, such as filaments and clusters, not in the voids. The Biermann's battery 
activated by ionization fronts~\citep{gnfezw00} generates fields that are too small in the low
density regions. Galactic outflows can in principle export high magnetic flux~\citep[e.g][]{bertone06}, 
but despite valuable efforts, the hydrodynamic transport that determines its
filling factor in the intergalactic space remains highly uncertain~\citep{rees06}.

\citet{Miniati2011ApJ...729...73M} proposed a mechanism to generate widespread intergalactic magnetic fields
that is associated with escape of high energy particles (heretofore CR for cosmic-ray) from 
the first generation of galaxies, also responsible for the re-ionization of the intergalactic medium (IGM).
Cosmic re-ionization is mostly driven by ionizing photons produced by massive OB
stars. When these stars come to an end, they explode as supernova whose
blast wave accelerates copious CR~\citep{krymsky77,axlesk77,bell78a,blos78}.
Owing to their much larger diffusive mean free path
compared to thermal particles, CR eventually escape from the parent galaxies into the 
IGM~\citep{Schlickeiser2002}. The charged CR protons carry an electric current in the IGM, $j_c$,
which is not balanced by the CR electrons, typically fewer in number by two 
orders of magnitude~\citep{Schlickeiser2002} and subject to mush more severe energy losses. 
Due to charge imbalance and induction effect a return current, $\vec j_t$,
carried by the cold thermal plasma is generated, which tends to cancel the CR
current, i.e. $j_t\simeq -j_c$. Due to the finite resistivity of the IGM plasma, an electric 
field, $\vec E=\eta \vec j_t$, is required to
draw the return current. This field opposes the propagation of the CR but is too weak
to prevent their escape. The plasma resistivity in an unmagnetized medium is determined
by its collisional rate which only depends on temperature, and can be written as
$\eta(T)=\eta_0 (T/{\rm K})^{-3/2}$ (Spitzer 1956), where $\eta_0$ is a constant.
Owing to the formation of structure in the universe, driven by gravitational instability,
considerable density and temperature inhomogeneities existed in the IGM long before 
the onset of re-ionization. Thus the electric field is also inhomogeneous and in particular
\begin{equation} \label{curle:eq}
\nabla\times\vec E=\nabla\times \eta \vec j_t \approx -\frac{3}{2}\eta \vec j_t\times\frac{\nabla T}{T} \ne 0!
\end{equation}
where we have neglected the term $\propto\nabla\times j_t$ and have used Spitzer's expression
for the resistivity. The important point is that, since the current is unrelated to the IGM 
inhomogeneities, the curl of the electric field is non vanishing.
The rotational component of the electric field sustains Faraday's induction and generates 
magnetic field.
The analysis of \cite{Miniati2011ApJ...729...73M} used the 
observed UV luminosity function of high redshift
galaxies to estimate the production rate of CR in those galaxies and the ensuing return
currents in the IGM, $j_t$. They also used simulations of structure formation to 
estimate the temperature gradient scale-length, $T/\nabla T$, entering the electric field
curl. They found that magnetic field is robustly generated
throughout intergalactic space at rate of 10$^{-17}-10^{-16}$
Gauss/Gyr, until the temperature of the intergalactic medium is
raised by cosmic re-ionization when the age of the universe is roughly $t=$1Gyr.
This value is consistent with the conservative values implied by gamma-ray measurements.

In this contribution, we present results from a cosmological numerical model that also 
includes the source terms associated with the resistive mechanism described above.
The numerical model and results are described below but  we anticipate that they validates
the quantitative findings of~\cite{Miniati2011ApJ...729...73M}. 
The rest of this manuscript is organized as follows.
The numerical model is described in Sec.~\ref{numod:sec}, the results are reported in Sec.~\ref{res:sec}, and Sec.~\ref{con:sec} contains summary and conclusions.

\section{Approximate Numerical Model of the Resistive Mechanism} \label{numod:sec}
We start from a numerical model of cosmological structure formation. This 
includes the evolution, in an expanding background space-time, 
of a collision-less dark matter fluid and a collisional baryonic gas, 
coupled through self-gravity, and subject to appropriate initial conditions. 
We then extend the hydrodynamics to include the effects of magnetic field in
the MHD approximation and modify the RHS of the governing equations
to include the various effects due to
the return current $j_t$ (see below). Because $j_t\simeq -j_c$, for convenience in the following we 
express the source terms associated to the return current in terms of $-j_c$.

The full set of equations solved in a cosmological model 
is given, e.g., in~\citet{mico07b}. Here we only describe the 
equations of hydrodynamics, which contain the modifications required by the resistive model.
As usual, these equations employ {\it comoving} 
density, $\rho$, and pressure, $p_g$, and {\it peculiar} velocity, ${\bf u}$. 
If $a(t)$ is the spatial scale factor of the 
universe and $\dot a$ its rate of change, density and pressure are modified as 
$\rho\leftarrow a^3\rho$, $p_g\leftarrow a^3P$ respectively, to scale out the effect of 
Hubble expansion. Likewise the peculiar velocity does not include the Hubble flow 
component~\cite[see, e.g.,][for details]{mico07b}.
For numerical reasons we find most convenient to rescale the
magnetic field as, ${\bf B}\leftarrow a^\frac{3}{2}{\bf B},~{\bf j_c}\leftarrow a^\frac{3}{2}{\bf j_c}$,
even though it does not scale out completely the adiabatic expansion effects.
So finally the governing equations read
\begin{eqnarray}
\label{rho:eq}
\frac{\partial\rho}{\partial t} + 
\frac{1}{a}\bf\nabla\cdot\left(\rho \bf u\right)&=&0,\\  
\label{mom:eq}
\frac{\partial\rho \bf u}{\partial t} + \frac{1}{a}\nabla\cdot
\left( \rho {\bf u} : {\bf u} + {p} -\frac{1}{4\pi}\bf B : \bf B\right)&=&
-\frac{\dot a}{a}\rho {\bf u}-\frac{1}{a}\rho \nabla\phi
-\frac{\bf j_c}{c} \times \bf B,\\
\label{ene:eq}
\frac{\partial\rho e}{\partial t} +\frac{1}{a}
\nabla\cdot\left[{\bf u} \left(\rho e+p\right)-\frac{1}{4\pi}\bf B (\bf B\cdot \bf u)\right] 
\!\!\!\!&=&\!\!\!\!
-2\frac{\dot a}{a}\rho e-\frac{1}{a}\rho {\bf u}\cdot\nabla\phi
- {\bf u} \cdot \left(\frac{\bf j_c}{c} \times \bf B\right)
+ \eta j_c^2.
\end{eqnarray}
Here $p=p_g+\frac{1}{8\pi}\bf B\cdot\bf B$ is
the total sum of the gas and magnetic pressures, $e=\half {\bf u\cdot
  u}+ e_{th}+\frac{1}{8\pi\rho}{\bf B\cdot B}$ is the total specific energy
density. The thermal energy is related to the pressure through a
$\gamma$-law equation of state, $e_{th}=p_g/\rho (\gamma-1)$, $\phi$
is the gravitational potential, ${\bf -j_c}$ the return current and
$\eta$ the resistivity.  The last term on the RHS of the momentum
equation describes the interaction between the magnetic field and the
return current, which also contributes the term before the last in the
energy equation. The last term in the energy equation represents
ohmic dissipation of the return current. The electric field, $\eta \bf
j_c$, driving this current does not appear in the momentum equation as
it is balanced by frictional forces, the same ones that are in fact
responsible for the dissipation.

The magnetic field evolution is described by Faraday's equation, with
the electric field given by Ohm's law.  The only electric fields are
those induced by motions of the magnetized fluid as well as the return
current, i.e. ${\bf E}=-({\bf u}/c)\times \bf B - \eta \bf j_c$. So the
modified induction equation (again in comoving quantities) reads
\begin{equation} \label{faraday:eq}
\frac{\partial \bf B}{\partial t} 
=\frac{1}{a}\nabla \times \left(\bf u\times \bf B\right)
+ \frac{1}{a}c \nabla\times  \eta {\bf j_c} - \frac{1}{2}\frac{\dot a}{a}\bf B.
\end{equation}
The full set of cosmological equations is solved with the code {\tt CHARM}~\citep{mico07b}. 
The MHD solver is based on the constrained-transport scheme and is 
described in details in~\cite{Miniati2011ApJS..195....5M}. The source terms associated with 
the resistive process have been implemented using a second order accurate time-centered
scheme.

Concerning the CR current, $j_c$, as discussed in~\cite{Miniati2011ApJ...729...73M}, 
it is typically dominated by the most luminous nearby galaxy.
For a bright $L^*$ galaxy at redshift $z\simeq6$, with star formation rate of 6.5 
$M_\odot$yr$^{-1}$~\cite{bouwens07,bouwens10}, assuming a Salpeter initial-mass-function 
and a 30\% conversion efficiency of supernova energy into CR, it is found
\begin{equation}
j_c\simeq  5.3\times 10^{-20}
\left(\frac{L}{L_*}\right)
\left(\frac{d}{\rm Mpc}\right)^{-2}
\left(\frac{p_{min}}{m_pc}\right)^{-0.3}
\left(1+\frac{p_{min}}{m_pc}\right)^{-1}
{\rm Amp\ m}^{-2} \label{jc:eq},
\end{equation}
where $d$ is the physical distance from the parent galaxy and 
$p_{min}$ is the minimum energy of the escaping CR.
In the following we consider a distance $d\sim 1$Mpc, which typically
separates bright $L^*$ galaxies at $z\sim6$,  and
a volume of linear size $L\ll d$ so that the distance dependence of the 
CR current can be neglected within the simulated volume.
In addition, the spatial inhomogeneities of $j_c$ are neglected, so $j_c$ is 
effectively a constant, whose value is set in accordance with Eq.~\ref{jc:eq}.
We neglect the effects due to the generated magnetic field on the CR propagation.
Though they introduce some degree of diffusivity in the CR propagation,
the net effect on the current should remain negligible for a field strength 
below 10$^{-15}$ G.

\section{Results}\label{res:sec}
\begin{figure} 
\begin{center}
\includegraphics[height=0.315\textheight, scale=1.0]{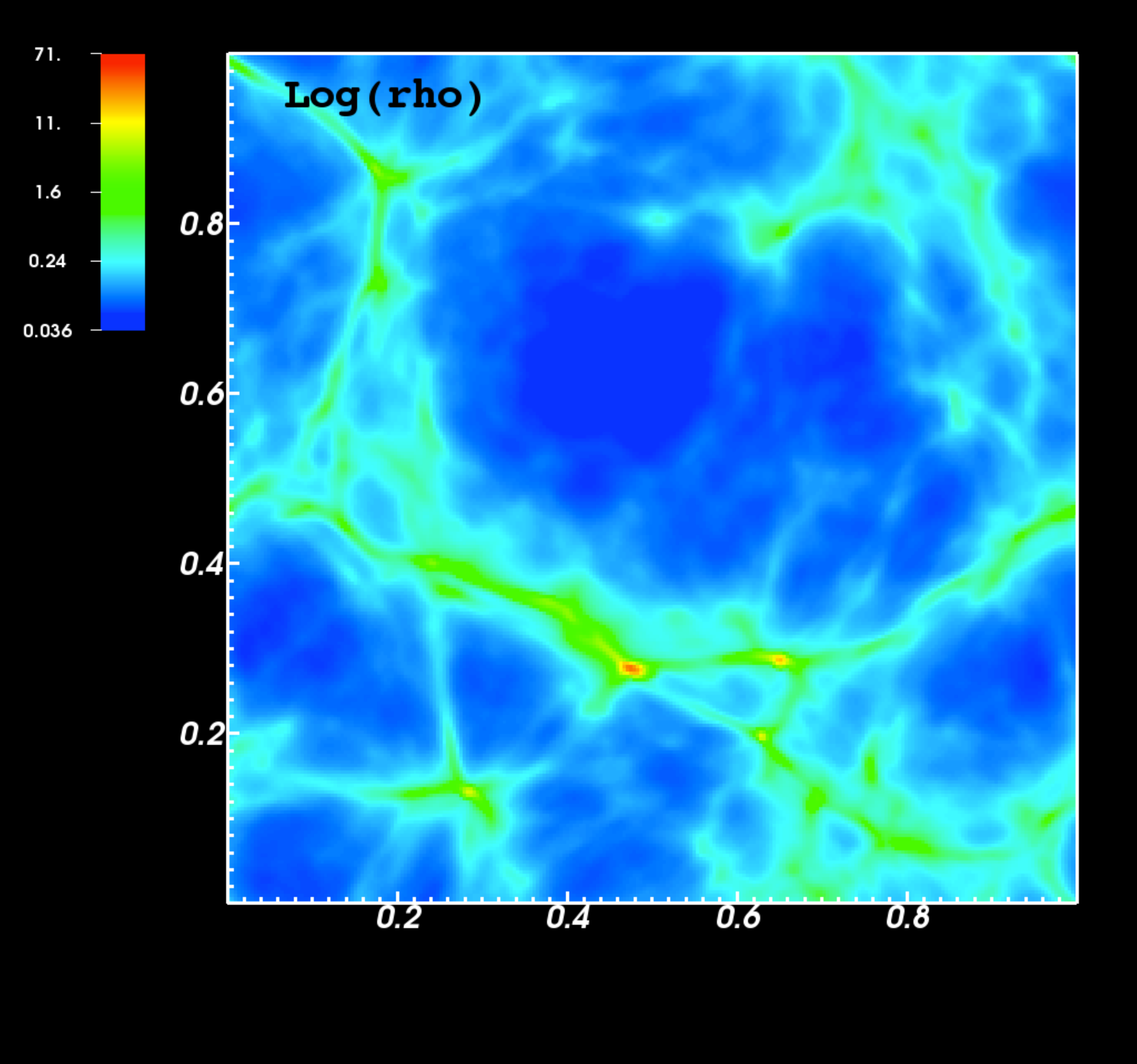}\includegraphics[height=0.315\textheight,scale=1.0]{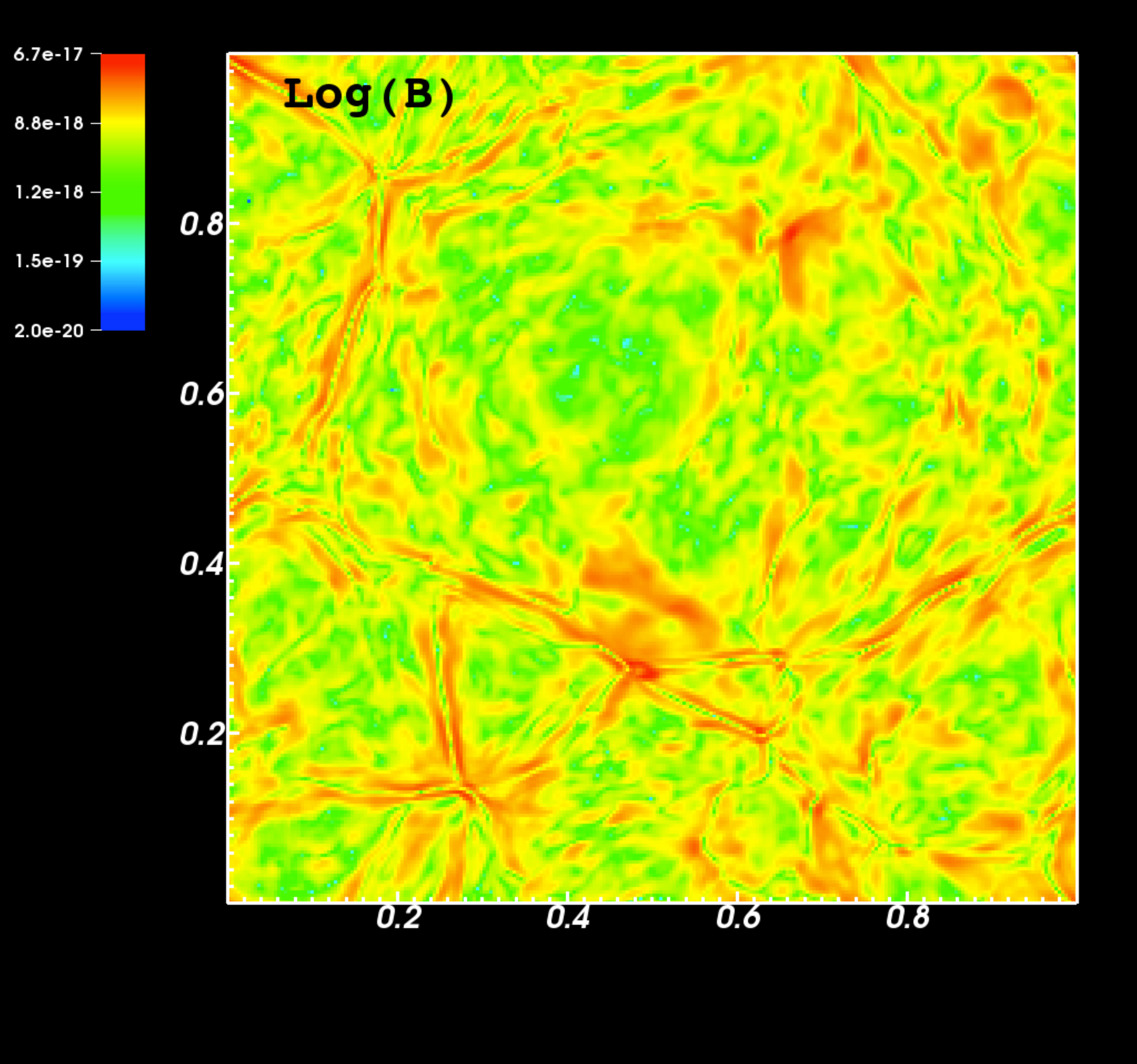}
\caption{Simulation snapshots for the distribution of baryonic gas (left) and magnetic field
strength (right) on a slice through the simulated volume at redshift $z=10$.
Gas density is in units of the average total (baryonic plus dark matter) matter density, 
and magnetic field is in Gauss. The physical length each panel's side is 12.7 kpc.
\label{igm:fig}}
\end{center}
\end{figure}
\begin{figure} 
\begin{center}
\includegraphics[height=0.315\textheight,scale=1.0]{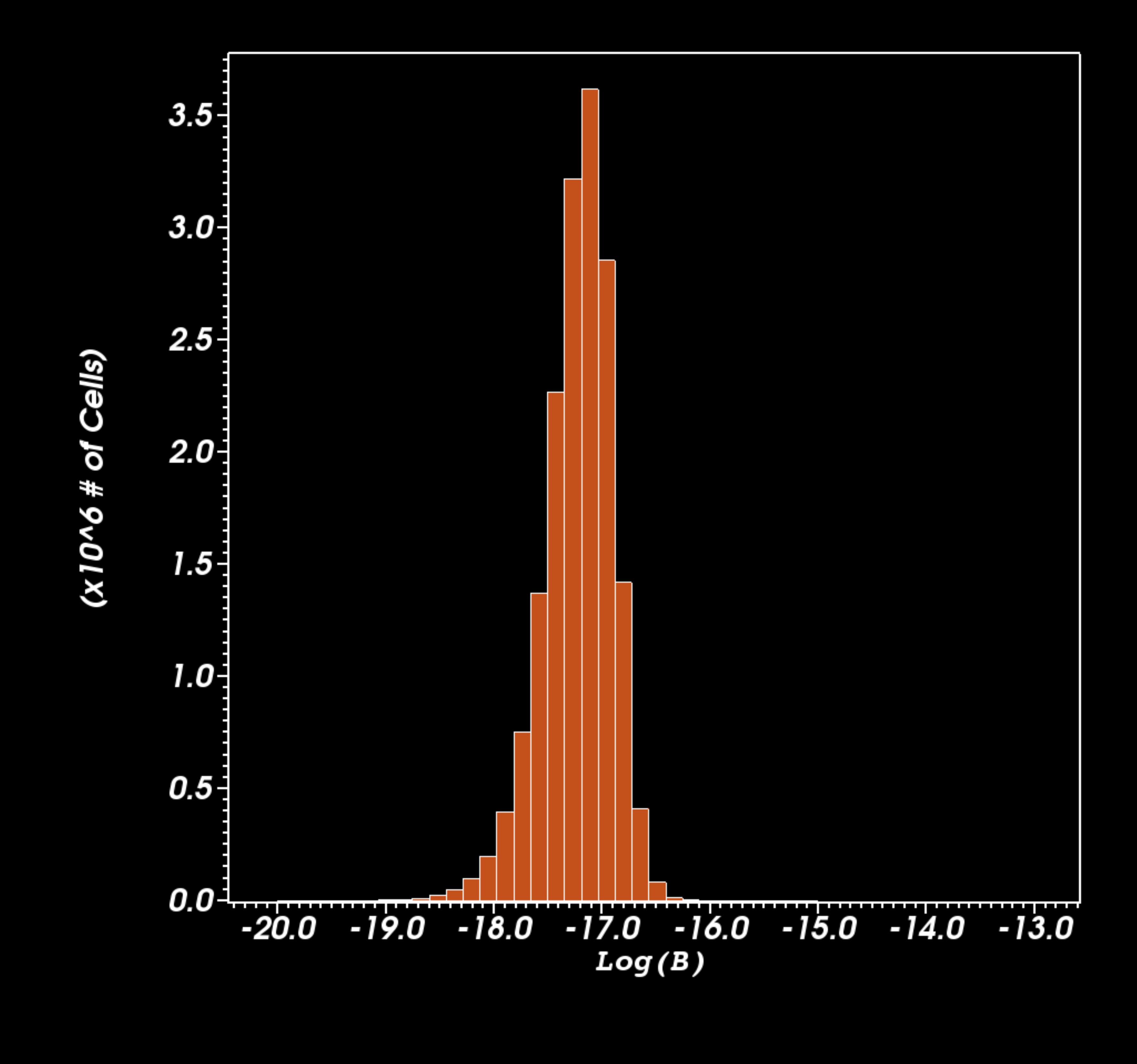}\includegraphics[height=0.315\textheight, scale=1.0]{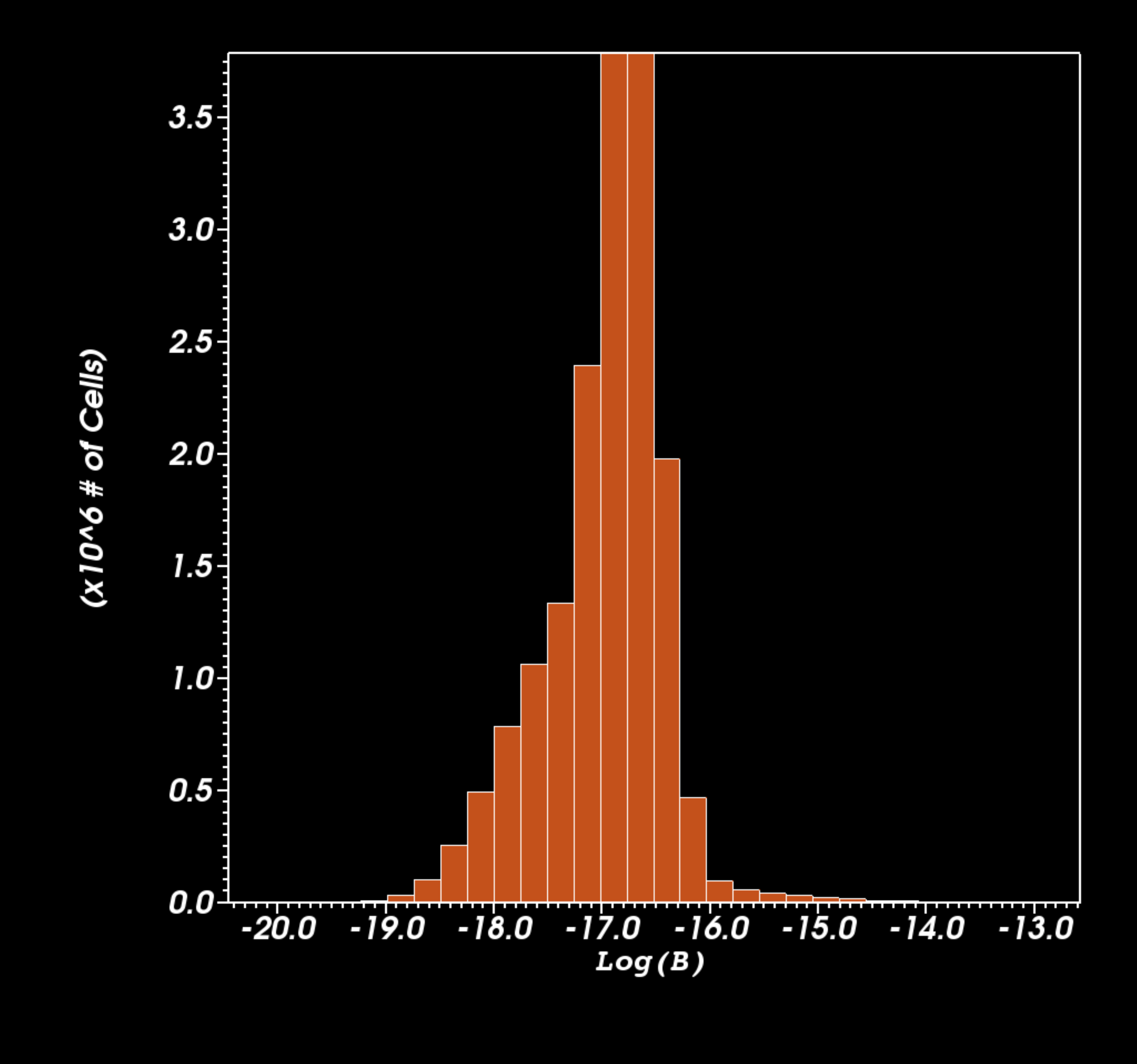}
\caption{Histrogram of magnetic field distribution as a function of its
strength at $z=10$ (left) and $z=6$ (right).
\label{bhist:fig}}
\end{center}
\end{figure}

Using the model described in the previous section we have run a model 
of cosmic structure formation that includes the resistive generation of magnetic 
field. For the purpose, we have adopted a concordance $\Lambda$-CDM 
based with cosmological parameters given by the results of the WMAP 
experiment~\cite{komatsuetal09}.
We use a computational box with 512$^3$ resolution elements and as many 
collision-less particles to describe the dark matter component.
For the physical size of the computational domain we consider two cases.
In the first one the simulated volume corresponds to a
cube of side $L=140$ comoving kpc.
At redshift $z\sim10$, corresponding to the epoch of the results presented below,
this translates into a physical size of 12.7 kpc and a spatial resolution of 25 pc,
sufficient to resolve temperature structures on physical scales of a few kpc, identified
in~\cite{Miniati2011ApJ...729...73M} as the most relevant for the resistive process.
Fig.~\ref{igm:fig} shows a snapshot of the distribution of the baryonic gas 
in units of the total matter density (left)
and magnetic field strength in Gauss (right), on a two-dimensional 
slice passing through the simulated volume  at redshift $z\sim 10$.
As anticipated above, the matter density distribution shows considerable structure,
which translates into a corresponding temperature structure due either to adiabatic
compression or shocks. Notice that the magnetic field ranges from 10$^{-18}$
to several times 10$^{-17}$ Gauss. An important feature is that the magnetic field 
with strength in this range fills is generated throughout the simulated volume.
This is better illustrated in the histogram on the left hand side of Fig.~\ref{bhist:fig},
which shows the distribution of magnetic field as a function of magnetic field 
strength. According to the histogram, all space is magnetized and the average 
magnetic field value is of order 10$^{-17}$ Gauss.

Given the small box size, at later time the formation of structure quickly 
saturates and the numerical model becomes inadequate to describe the intergalactic
medium. We have thus run a second model which employs a box ten times as large as for 
the previous case, $L=1.4$ comoving Mpc. We run this model down to redshift $z\sim 6$,
when cosmic re-ionization raises the IGM temperature, effectively shutting down the 
resistive mechanism.
Since we employ the same numerical resources as before, so the resolution is at least 
390 kpc physical, which is still sufficient to resolve the desired temperature structures.
The histogram on right hand side of Fig.~\ref{bhist:fig} shows the distribution of the magnetic
field strength in this larger simulated volume at redshift $z\sim 6$. It shows that 
the magnetic field continues to grow in strength from redshift 10 to 6 and at this later
time it fills the intergalactic volume with a strength between 10$^{-18}$ and 10$^{-16}$.
The histogram peaks at a value of several times 10$^{-17}$, which is quantitative consistent
with the estimates in~\cite{Miniati2011ApJ...729...73M}.

Finally, although the governing equations~\ref{rho:eq}-\ref{faraday:eq} include 
the ingredients necessary to drive non resonant amplification of magnetic 
field~\citep{bell04,bell05}, there is no appreciable effect is our simulations.
This is because for the tiny resistive magnetic field the growth timescales of the 
non-resonant mechanism is too slow on the spatial scales resolved by our simulation.
It remains unclear whether or not this mechanism may efficiently kick in at later times.

\section{Summary and Conclusions}\label{con:sec}
\citet{Miniati2011ApJ...729...73M} proposed a mechanism for the generation of magnetic 
seeds that is based the finite resistivity of the low temperature IGM in the high redshift universe.
In that model, CR escaping from the first generation of galaxies, induce 
a return current and an electric field in the IGM plasma. The electric
field, $\vec E=\eta(T)\vec j$, depends on the IGM temperature, which 
is highly inhomogeneous due to adiabatic 
contraction and shocks produced by structure formation.
It is easy to show that this electric field has a non vanishing
curl which sustains the growth of magnetic field.
The analytic results in \citet{Miniati2011ApJ...729...73M}  were
consistent with conservative values of magnetic fields in voids 
inferred cosmic-voids by gamma-ray experiments.
Here we have implemented the source terms 
describing the resistive process in the cosmological code {\tt CHARM}.
To keep the problem tractable we made a number of reasonable approximations.
Nevertheless, the numerical model allows a more self-consistent estimate of the magnetic field generation by the resistive process. We have carried out numerical experiments
using two different computational boxes to best estimate the generated magnetic fields at 
different cosmic epochs of interest.
In conclusion the numerical estimates substantiate the earlier analysis 
in~\cite{Miniati2011ApJ...729...73M}, showing that most of the made approximations and 
assumptions  were essentially correct.

\acknowledgements
The numerical work presented here was carried out on Brutus HP cluster facility hosted at the 
department of computer science of ETH-Z\"urich. The research leading to these results has 
received funding from the European Research Council under the European
Community's Seventh Framework Programme (FP7/2007-2013) / ERC grant agreement no. 247039.

\bibliographystyle{mn2e.bst}
\bibliography{../biblio/books,../biblio/codes,../biblio/papers,../biblio/proceed,../biblio/papigm}

\end{document}